\documentclass[aps,prb,twocolumn,superscriptaddress]{revtex4}
\usepackage{amsfonts}
\usepackage{amssymb}
\usepackage{amsmath}
\usepackage{graphicx}
\usepackage{epsfig}
\usepackage{color}
\usepackage{amsmath,bm}

\begin{document}

\title{Dynamical preparation of a steady ODLRO state in the Hubbard model
with local non-Hermitian impurity}
\author{X. Z. Zhang}
\affiliation{College of Physics and Materials Science, Tianjin Normal University, Tianjin
300387, China}
\author{Z. Song}
\email{songtc@nankai.edu.cn}
\affiliation{School of Physics, Nankai University, Tianjin 300071, China}

\begin{abstract}
The cooperation between non-Hermiticity and interaction brings about a lot
of counterintuitive behaviors, which are impossible to exist in the
framework of the Hermitian system. We study the effect of a non-Hermitian
impurity on the Hubbard model in the context of $\eta $ symmetry. We show
that the non-Hermitian Hubbard Hamiltonian can respect a full real spectrum
even if a local non-Hermitian impurity is applied to. The balance between
dissipation of single fermion and on-site pair fluctuation results in a
highest-order coalescing state with off-diagonal long-range order (ODLRO).
Based on the characteristic of High-order EP, the critical non-Hermitian
Hubbard model allows the generation of such a steady superconducting-like
state through the time evolution from an arbitrary initial state, including
the vacuum state. Remarkably, this dynamic scheme is insensitive to the
on-site interaction and entirely independent of the locations of particle
dissipation and pair fluctuation. Our results lay the groundwork for the
dynamical generation of a steady ODLRO state through the critical
non-Hermitian strongly correlated system.
\end{abstract}

\maketitle

%\affiliation{College of Physics and Materials Science, Tianjin Normal University, Tianjin
%300387, China}

%\affiliation{School of Physics, Nankai University, Tianjin 300071, China}

%\email{zhangxz@tjnu.edu.cn}

\section{Introduction}

\label{introduction} In recent years, nonequilibrium dynamics induced by
dissipation exhibits intriguing properties. As an effective description, the
non-Hermitian Hamiltonian arises when the system experiences dissipation to
an environment \cite{Daley2014,Dalibard1992}. Recent years have seen a
growing interest in non-Hermitian descriptions of condensed-matter systems
which have not only extended the domain of condensed-matter physics with
inspiring insights \cite%
{Lee2016,Kunst2018,Yao2018,Gong2018,El-Ganainy2018,Nakagawa2018,Shen2018,Wu2019,Yamamoto2019,Song2019,Yang2019,Hamazaki2019,Kawabata2019,Kawabata2019a,Lee2019,Yokomizo2019,Jin2020}
but also provided a fruitful framework to elucidate inelastic collisions
\cite{Xu2017}, disorder effects \cite{Shen2018,Hamazaki2019}, and
system-environment couplings \cite{Nakagawa2018,Yang2019,Song2019}. In
particular, the interplay between non-Hermiticity and interaction can give
rise to exotic quantum many-body effect, ranging from non-Hermitian
extensions of Kondo effect \cite{Nakagawa2018,Lourenfmmodeboxclsecio2018},
many-body localization \cite{Hamazaki2019}, Fermi surface in coordinate
space \cite{Mu2020}, to fermionic superfluidity \cite{Yamamoto2019,Okuma2019}%
. It has been shown that the cooperation between the non-Hermiticiy and
interaction can alter drastically the macroscopic behavior that has been
established in the Hermitian physics. Meanwhile, the generation of the
superconducting-like states that have been induced by dissipation has
received great attention \cite%
{Diehl2008,Kraus2008,Sentef2016,Mitrano2016,Coulthard2017}. However, these
schemes rely on judicious engineering of the system parameters to avoid
thermalization.

Exceptional points (EPs) are degeneracies of non-Hermitian operators where
the corresponding eigenstates coalesce into one state leading to a
incomplete Hilbert space \cite{Berry2004,Heiss2012,Miri2019,Zhang2020}.%
% Classical and quantum
%photonic systems with EPs have attracted tremendous attention due to the
%dramatically change in the vicinity of EPs.
The peculiar features around EP have sparked tremendous attention to the
classical and quantum photonic systems. The corresponding intriguing
dynamical phenomena include asymmetric mode switching \cite{Doppler2016},
topological energy transfer \cite{Xu2016}, robust wireless power transfer
\cite{Assawaworrarit2017}, and enhanced sensitivity \cite%
{Wiersig2014,Wiersig2016,Hodaei2017,Chen2017} depending on their EP
degeneracies. Notably, the high-order EP with $N$ coalescent states (EPN)
attract much more interest recently. Many works have been devoted to the
formation of the EPN and corresponding topological characterization in both
theoretical and experimental aspects \cite%
{Ding2016,Xiao2019,Pan2019,Zhang2020a}. %at which more than two eigenstates
%coalesce have received a lot of attention due to their topological property
%and unique dynamical property.

Given the above two rapidly growing fields, namely, the non-Hermitian
interacting system and dynamics of the EP, we are motivated to examine how
non-Hermiticity impacts on the interacting system, especially the dynamics
when the high-order EPs of the interacting system presents. In this paper,
we uncover the effect of cooperation between particle dissipation and pair
fluctuation on the strongly-correlated system by concentrating on the
non-Hermitian Hubbard model. We show that the considered interacting system
can respect the full real spectrum even no obvious symmetry presents.
Remarkably, we find that a balance local non-Hermitian impurity can induce
the formation of high-order EPs in the spectrum in the way that degenerate
states with different symmetry of parent Hermitian system coalesce. The $%
\eta $-pairing mechanism plays a vital role to achieve this intriguing
property. Based on the performance of the system at EP, a scheme that
produces a nonequilibrium steady superconducting-like state is proposed.
Specifically, for an arbitrary initial state even a vacuum state, the
critical system can drive it to the coalescent state that favors
superconductivity manifested by the off-diagonal long-range order (ODLRO).
Such a dynamical scheme can be realized no matter where the local coupling
is applied and the correlation of the final steady state is independent of
the relative distance between the two sites. Therefore, our finding is
distinct from the previous investigations \cite{Tindall2019,Kaneko2019}, and
offer an alternative mechanism for generating superconductivity through
nonequilibrium dynamics. On the other hand, the remarkable observation from
our work can trigger further studies of both fundamental aspects and
potential applications of critical non-Hermitian strongly correlated systems.

The rest of this paper is organized as followed. In Section ~\ref{model}, we
present the general property of the considered non-Hermitian Hamiltonian
especially focuses on the mechanism of the formation of high-order EP.
Section ~\ref{local} is devoted to demonstrating how a local coupling can
make the degenerate eigenstate coalesce. In Section ~\ref{dynamics}, the
generation of the superconducting state based on the critical non-Hermitian
Hubbard model is proposed. Section ~\ref{summary} concludes this paper. Some
nonessential details of our calculation are placed in the Appendix.

\section{Model}

\label{model} We consider a non-Hermitian extension of Hubbard model for
two-component fermions. The Hamiltonian is in the form
\begin{eqnarray}
H &=&H_{0}+H_{I},  \label{H} \\
H_{0} &=&-\sum_{<i,j>}\sum_{\sigma =\uparrow ,\downarrow }t_{ij}c_{i,\sigma
}^{\dagger }c_{j,\sigma }+\text{\textrm{H.c.}}  \notag \\
&&+U\sum_{j=1}^{2N}\left( n_{j,\uparrow }-\frac{1}{2}\right) \left(
n_{j,\downarrow }-\frac{1}{2}\right) ,  \label{H0} \\
H_{I} &=&\sum_{j}\bm{h}_{j}\cdot \bm{\eta }_{j},  \label{HI}
\end{eqnarray}%
where $\bm{h}_{j}=g_{j}\left( \lambda ,\text{ }0,\text{ }i\gamma \right) $
with $\left\{ {g_{j}}\right\} $ describing inhomogeneity and a set of
arbitrary numbers, and the corresponding $\eta $-operators are defined by
\begin{eqnarray}
\eta _{j}^{+} &=&\eta _{j}^{x}+i\eta _{j}^{y}=\left( -1\right)
^{j}c_{j,\uparrow }^{\dagger }c_{j,\downarrow }^{\dagger },  \label{eta1} \\
\eta _{j}^{-} &=&\eta _{j}^{x}-i\eta _{j}^{y}=\left( -1\right)
^{j}c_{j,\downarrow }c_{j,\uparrow },  \label{eta2} \\
\eta _{j}^{z} &=&\frac{1}{2}\left( n_{j,\uparrow }+n_{j,\downarrow
}-1\right) ,  \label{eta3}
\end{eqnarray}%
obeying the Lie algebra, i.e., $[\eta _{j}^{+},$ $\eta _{j}^{-}]=2\eta
_{j}^{z}$ and $[\eta _{j}^{z},$ $\eta _{j}^{\pm }]=\pm \eta _{j}^{\pm }$.
Here $c_{i,\sigma }$ is the annihilation (creation) operator for an electron
at site $i$ with spin $\sigma $ and $n_{i,\sigma }=c_{i,\sigma }^{\dagger
}c_{i,\sigma }$. $H_{0}$ is a standard Hubbard model on a bipartite lattice
where $t_{ij}$ and $U$ play the role of kinetic and interaction energy
scale; $H_{I}$ describes the non-Hermitian impurity that consists of on-site
pair fluctuation and imaginary magnetic field, which can be achieved by
coupling the system to the environment and are within the reach of both the
ultracold atom \cite{Lee2014} and photonic experiments \cite%
{Fausti2011,Hu2014,Kaiser2014,Mitrano2016,Cantaluppi2018}.

Now we turn to investigate the symmetry of the considered model. The
Hamiltonian in Eq. (\ref{H0}) has a rich structure due to the two sets of
SU(2) symmetries it possesses. The first, often referred to as $\eta $
symmetry, is central to this paper and can be characterized by the
generators $\eta ^{\pm }=\sum_{j}\eta _{j}^{\pm }$ and $\eta
^{z}=\sum_{j}\eta _{j}^{z}$. It relates to spinless quasiparticles (doublons
and holons) and can be interpreted as a type of particle-hole symmetry. The
second of these is spin symmetry, and the corresponding generator $\bm{s}$
can be obtained by replacing $c_{j,\downarrow }$ of Eqs. (\ref{eta1})-(\ref%
{eta3}) with $\left( -1\right) ^{j}c_{j,\downarrow }^{\dagger }$, that is
\begin{eqnarray}
s^{+} &=&s^{x}+is^{y}=\sum_{j}c_{j,\uparrow }^{\dagger }c_{j,\downarrow }, \\
s^{-} &=&s^{x}-is^{y}=\sum_{j}c_{j,\downarrow }^{\dagger }c_{j,\uparrow }, \\
s^{z} &=&\frac{1}{2}\sum_{j}\left( n_{j,\uparrow }-n_{j,\downarrow }\right).
\end{eqnarray}
It can be readily seen that the operators in Eqs. (\ref{eta1})-(\ref{eta3})
fulfill the relations $\left[ H,\eta ^{\pm }\right] =\left[ H,\eta ^{z}%
\right] =0$ and commute with all the generators of the spin symmetry. The
presence of $H_{I}$ spoils two such SU(2) symmetries. However, the two
Hamiltonians $H_{0}$ and $H_{I}$ commute with each other when the
interacting strength $g_{j}$ is homogeneous such that $H_{I}$ can be treated
as $g\left( \lambda \eta ^{x}+i\gamma \eta ^{z}\right) $. As such the two
Hamiltonians shares the common eigenstates. Although the two Hamiltonians
commutes with each other, $H_{I}$ profoundly changes the energy spectrum of
the whole system due to the emergence of EPN, which is distinct from the
Hermitian case. In this paper, we focus on the subspace spanned by the $\eta
$-pairing state $\left\vert \psi _{N_{\eta }}\right\rangle =\Omega
^{-1}\left( \eta ^{+}\right) ^{N_{\eta }}\left\vert \mathrm{Vac}%
\right\rangle $, where $\left\vert \mathrm{Vac}\right\rangle $ is a vacuum
state with no electrons and $N_{\eta }$ is the number of $\eta $ pairs. What
makes the state $\left\vert \psi _{N_{\eta }}\right\rangle $ is special is
the fact that it has been shown to have off-diagonal long-range order
(ODLRO) in the form of doublon-doublon correlations, $\langle \psi _{N_{\eta
}}|\eta _{i}^{\dagger }\eta _{j}^{-}\left\vert \psi _{N_{\eta
}}\right\rangle =\mathrm{const}$, ($i\neq j$). The nonzero value of such
quantity implies both the Meissner effect and flux quantization and
therefore provides a possible definition of superconductivity \cite{Yang1962}%
. These states are the $2N+1$ fold degenerate eigenstates of both $H_{0}$
and $\eta ^{2}$. \cite{Yang1989} Hence, all these states can be expressed as
$\{\left\vert N,l\right\rangle \}$, where $N$ and $l$ are associated with
the eigenvalues of $\eta ^{2}$ and $\eta ^{z}$, i.e.,
\begin{eqnarray}
\eta ^{2}\left\vert N,l\right\rangle &=&N\left( N+1\right) \left\vert
N,l\right\rangle , \\
\eta ^{z}\left\vert N,l\right\rangle &=&l\left\vert N,l\right\rangle ,\text{
}l\in \left[ -N,N\right] ,
\end{eqnarray}%
with $\left\vert N,l\right\rangle =\Omega ^{-1}\left( \eta ^{+}\right)
^{N+l}\left\vert \mathrm{Vac}\right\rangle $ and $\Omega =\sqrt{C_{2N}^{N+l}}
$. For the homogeneous $g_{j}=g$, the matrix of $H$ in the doublon invariant
subspace has the form%
\begin{eqnarray}
H_{\mathrm{doub}} &=&\frac{\lambda g}{2}\sum_{l=-N}^{N-1}J_{l}\left\vert
N,l\right\rangle \left\langle N,l+1\right\vert +\text{\textrm{H.c.}}  \notag
\\
&&+\sum_{l=-N}^{N}(\frac{NU}{2}+i\gamma gl)\left\vert N,l\right\rangle
\left\langle N,l\right\vert .
\end{eqnarray}%
with $J_{l}=\sqrt{(N-l)(N+l+1)}$. It describes a $\mathcal{PT}$ -symmetric
hypercube graph of $2N+1$ dimension \cite{Zhang2012}. The EP at $\left\vert
\lambda \right\vert =\left\vert \gamma \right\vert $ divides the system into
two different phases: $\mathcal{PT}$ -symmetric broken ($\left\vert \lambda
\right\vert <\left\vert \gamma \right\vert $) and unbroken regions ($%
\left\vert \lambda \right\vert >\left\vert \gamma \right\vert $). The unique
feature of $H_{\mathrm{doub}}$ is that all the eigenstates coalesce at $%
\left\vert \lambda \right\vert =\left\vert \gamma \right\vert $ forming a
high-order EP with order of $2N+1$. The corresponding coalescent state is
\begin{equation}
\left\vert \Phi _{\mathrm{c}}\right\rangle =(\frac{i}{2})^{N}\sum_{l=-N}^{N}%
\sqrt{C_{2N}^{N+l}}\left( i\right) ^{l}\left\vert N,l\right\rangle ,
\label{coa}
\end{equation}%
with Dirac normalization. If the system is initialized in this invariant
subspace, then all the dynamical property of the system is solely determined
by the non-Hermitian Hamiltonian $H_{\mathrm{doub}}$. Specifically, when $H_{%
\mathrm{doub}}$ respects full real spectrum, the physical observable will
show a periodic oscillating behavior. This dynamic behavior will be
significantly changed when the critical non-Hermitian impurity $\lambda
g\left( \eta _{x}+i\gamma \eta _{z}\right) $ is applied, that is, any
initial state will be forced to evolve to the coalescent state. The
corresponding physical observable reflects the property of such a state.
This mechanism will be served as the building block to investigate the ODLRO
of the nonequilibrium steady state.

\section{Local dissipation and pair fluctuation}

\label{local} In this section, we will demonstrate how does the local
impurity can dramatically change the structure of the system. To proceed, we
consider $H_{I}$ of Eq. (\ref{HI}) in which the local impurities are applied
to a part of lattice sites. An extreme case is that the local dissipation is
applied to one lattice site. Compared with the homogeneous case, the $\eta $%
-pairing states $\{\left\vert N,l\right\rangle \}$ cannot form an invariant
subspace due to the relation of $\left[ H_{I},\text{ }H_{0}\right] \neq 0$.
Consequently, we cannot judge the EP of the whole system only by the
performance of $H_{I}$. However, we can infer the property of $H$ by its
counterpart $\overline{H}$ that can be obtained by the transformation $%
\mathcal{S=}$ $e^{i\theta \eta ^{y}}$ with $\tan ^{-1}\theta =i\gamma
/\lambda $. It represents a rotation in the $\eta ^{x}$-$\eta ^{z}$ plane
around $\eta ^{y}$ axis by an complex angle $\theta $ determined by the
strength of local pair fluctuation and imaginary field. Notably, the
rotation operator is valid at arbitrary $\gamma /\lambda $ unless at the EP (%
$\gamma /\lambda =\pm 1$) of $H_{I}$, where $\bm{h}_{j}\cdot \bm{\eta }_{j}$
cannot be diagonalized. Applying the transformation $\mathcal{S}$, $H$ is
transformed to $\overline{H}=H_{0}+\sqrt{\lambda ^{2}-\gamma ^{2}}\sum_{ j
}g_{j}\zeta _{j}^{x}$ where the new set of operators $\zeta _{j}^{\pm ,z}=%
\mathcal{S}^{-1}\eta _{j}^{\pm ,z}\mathcal{S}$ also obey the Lie algebra,
that is $[\zeta _{j}^{+},$ $\zeta _{j}^{-}]=2\zeta _{j}^{z}$ and $[\zeta
_{j}^{z},$ $\zeta _{j}^{\pm }]=\pm \zeta _{j}^{\pm }$. Notice that $\zeta
_{j}^{\pm ,z}\neq (\zeta _{j}^{\mp ,z})^{\dagger }$ owing to the complex
angle $\theta $. However, the matrix form of $\overline{H}$ is Hermitian
under the biorthogonal basis of $\{\mathcal{S}^{-1}\left\vert \phi
\right\rangle \}$ and $\{\mathcal{S}^{\dagger }\left\vert \phi \right\rangle
\}$ when $\left\vert \lambda \right\vert >\left\vert \gamma \right\vert $,
where $\left\{ \left\vert \phi \right\rangle \right\} $ represents all the
eigenstates of $\eta ^{z}$. It is worth pointing out that such a
transformation solely depends on the ratio of $\gamma /\lambda $ and hence
the spectrum is entirely real, even though a nonzero inhomogeneous $g_{j}$
presents. Furthermore, it also indicates that the presence of the local pair
fluctuation and imaginary field breaks the SU(2) symmetry of the system but
remains the entirely real spectrum without symmetry protection.
\begin{figure*}[tbp]
\centering
\includegraphics[width=0.7\textwidth]{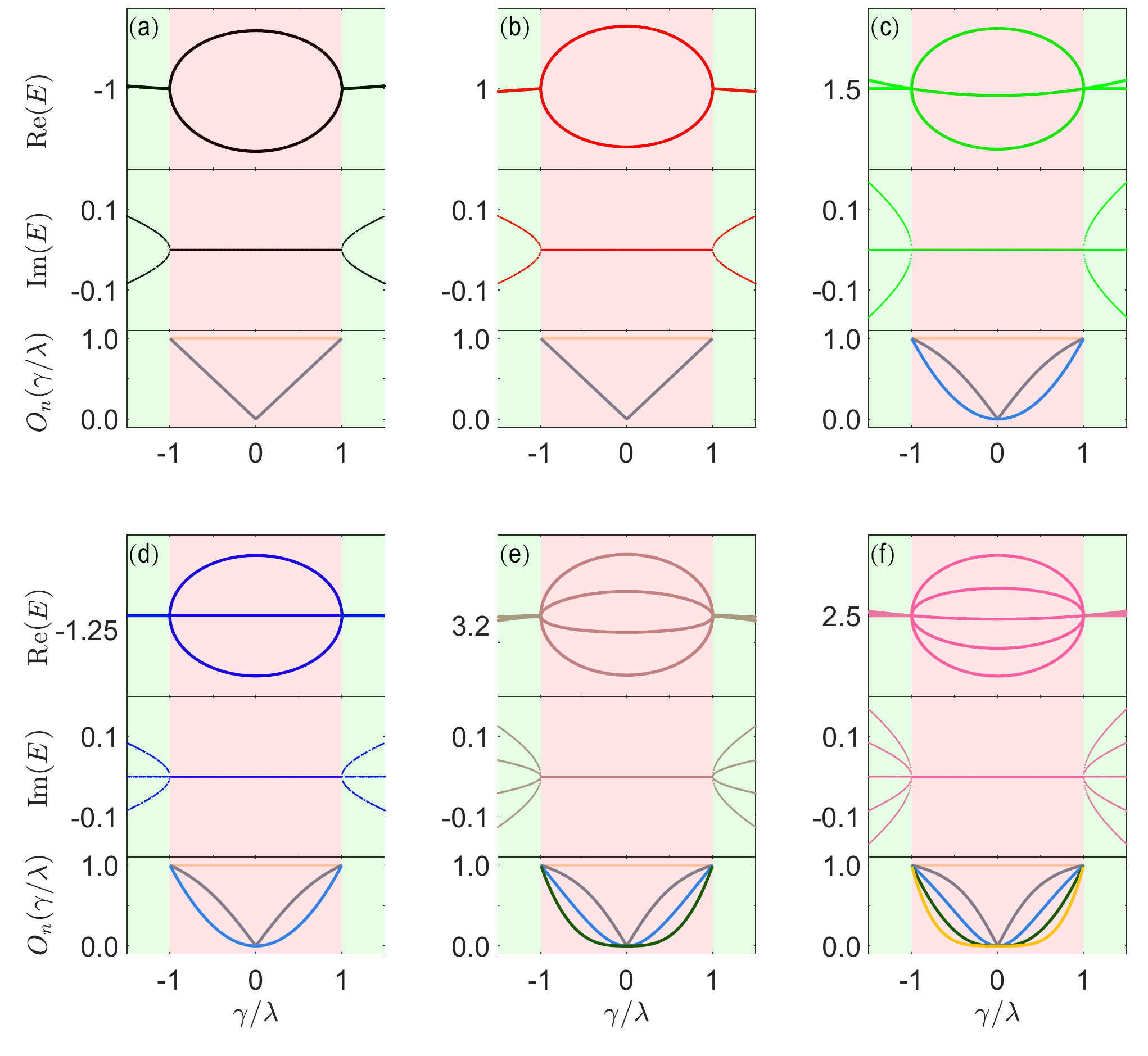}
\caption{Plots of the eigenenergies $E$ of $H$ and $O_{n}\left( \protect%
\lambda /\protect\gamma \right) $ as functions of $\protect\lambda /\protect%
\gamma $ for the system with one site subjected to the local imaginary field
for (a-c) ($N=1$) and two sites subjected to the local imaginary field for
(d-f) ($N=2$). The other system parameters are $U=3t$, $g_{1}=0.2t$ ($%
g_{j}=0 $, $j\neq 1$) for (a) and $U=2.5 $, $g_{1}=0.2t$, $g_{2}=0.1t$ ($%
g_{j}=0$, $j\neq 1,2$). The red and green shaded regions divides the system
into two phases, namely the phase with full real spectrum and complex
spectrum. The $2 $-site non-Hermitian Hubbard model contains four EPs with
order of $2$ and one EP with order of $3$. Figs. \protect\ref{fig_spectrum}%
(a)-\protect\ref{fig_spectrum}(c) covers all the possible types of EP. The
corresponding coalescent can be identified by $\protect\eta$ and spin
symmetry. Figs. \protect\ref{fig_spectrum}(a)-\protect\ref{fig_spectrum}(b)
represent the EP with same order which is formed in the subspace denoted by $%
\protect\eta ^{2}=3/4$, $s_{z}=1/2$ and $\protect\eta ^{2}=3/4$, $s_{z}=-1/2$%
, respectively. The Fig. \protect\ref{fig_spectrum}(c) depicts the formation
of EP3 within the subspace with $\protect\eta ^{2}=2$, $s_{z}=0$. Figs.
\protect\ref{fig_spectrum}(d)-\protect\ref{fig_spectrum}(f) describes three
types of EP in the subspace of $\protect\eta ^{2}=3/4$, $s_{z}=1$, $\protect%
\eta ^{2}=3/4$, $s_{z}=1/2$, and $\protect\eta ^{2}=6$, $s_{z}=0$,
respectively. It is shown that a local non-Hermitian impurity can induce
coalescent of the eigenstates which can be demonstrated by the behavior of
the level repulsion around $\left\vert \protect\lambda /\protect\gamma %
\right\vert =1$ in the upper two panels. The lower panel of each subfigures
shows the degree of similarity among eigenstates. Notice that all the
coalescent states in each subspace possess the geometric multiplicity of $1$%
, which indicates that the formation of such states shares the same
mechanism.}
\label{fig_spectrum}
\end{figure*}

Now we investigate the effect of local term on the $\eta $-pairing subspace.
The Hermiticity of the matrix form of $\overline{H}$ guarantees the validity
of the approximation methods in quantum mechanics. When $\left\vert \gamma
\right\vert \rightarrow \left\vert \lambda \right\vert $, the local term can
be treated as a perturbation. With the spirit of the degenerate perturbation
theory, the eigenvalues up to the first order are determined by the matrix
form of $\sqrt{\lambda ^{2}-\gamma ^{2}}\sum_{j}g_{j}\zeta _{j}^{x}$ in the
subspace spanned by $\{\left\vert N,l\right\rangle ^{\prime }\}$ with $%
\left\vert N,l\right\rangle ^{\prime }=\mathcal{S}^{-1}\left\vert
N,l\right\rangle $. The corresponding perturbed matrix is referred to as $%
\overline{H}_{\mathrm{doub}}^{\prime }$, whose elements $\overline{H}_{%
\mathrm{doub}}^{\prime }$ are given as $\langle \overline{N,l^{\prime }}|%
\overline{H}_{\mathrm{doub}}^{\prime }\left\vert N,l\right\rangle ^{\prime
}=\left( \mathcal{\delta }_{l,l^{\prime }+1}+\delta _{l+1,l^{\prime
}}\right) G\sqrt{\lambda ^{2}-\gamma ^{2}}J_{l}/4N$, where $G=\sum_{j}g_{j}$
and $1/2N$ stems from the translation symmetry of the $\eta $-pairing state.
$\{\langle \overline{N,l}|\}$ are biorthogonal left eigenvector in the form
of $\{\langle N,l|\mathcal{S}\}$. On the other hand, $H_{I}$ inevitably
induce the tunneling from the considered subspace to the other subspace such
that the high-order correction to the eigenenergies should be considered.
However, the high-order correction term is proportional to the $n$-th power
of $\left( \lambda ^{2}-\gamma ^{2}\right) /U$. If the large $U$ limit and
the condition of $\left\vert \gamma \right\vert \rightarrow \left\vert
\lambda \right\vert $ are taken, then one can throw safely high-order
correction. Performing the inverse transformation $\mathcal{S}\overline{H}_{%
\mathrm{doub}}^{\prime }\mathcal{S}^{-1}$, one can obtain the effective
Hamiltonian in the doublon invariant subspace%
\begin{eqnarray}
H_{\mathrm{doub}} &=&\frac{\lambda G}{4N}\sum_{l=-N}^{N-1}J_{l}\left\vert
N,l\right\rangle \left\langle N,l+1\right\vert +\text{\textrm{H.c.}}  \notag
\\
&&+\sum_{l=-N}^{N}(\frac{i\gamma lG}{2N}+\frac{NU}{2})\left\vert
N,l\right\rangle \left\langle N,l\right\vert .  \label{H_eff}
\end{eqnarray}%
It is a typical non-Hermitian hypercube. When $\left\vert \lambda
\right\vert =\left\vert \gamma \right\vert $, an EP($2N+1$) will be formed
no matter which site a local imaginary field is applied to. The
corresponding coalescent state is the same with Eq. (\ref{coa}). We stress
that such coalescent state is protected by pair interaction during the time
evolution.

To verify the conclusion above, we plot the eigenenergies of $H$ as
functions of $\gamma /\lambda $ in Fig. \ref{fig_spectrum}. Here, we assume
that the local imaginary field and pair fluctuation is applied to site $1$
unless stated otherwise. $O_{n}\left( \gamma /\lambda \right) $ is
introduced to quantify the similarity between eigenstates $\left\vert \Phi
_{1}\left( \gamma /\lambda \right) \right\rangle $ and $\left\vert \Phi
_{n}\left( \gamma /\lambda \right) \right\rangle $ of $H_{\mathrm{doub}}$,
which is defined as
\begin{equation}
O_{n}\left( \gamma /\lambda \right) =\left\vert \langle \Phi _{1}\left\vert
\Phi _{n}\right\rangle \right\vert /\sqrt{\left\vert \langle \Phi
_{1}\left\vert \Phi _{1}\right\rangle \right\vert \left\vert \langle \Phi
_{n}\left\vert \Phi _{n}\right\rangle \right\vert },
\end{equation}%
where $\left\vert \Phi _{1}\right\rangle $ is the eigenstate with lowest
eigenenergy. From the Fig. \ref{fig_spectrum}, we find that a critical
imaginary field cannot only make the degenerate $\eta $-pairing states of $%
H_{0}$ coalesce but also turn the other degenerate states to a coalescent
state and therefore form the multiple high-order EPs of the spectrum. This
phenomenon can be understood as follow: for the other eigenstate $\left\vert
\varphi _{0}\right\rangle $ of $H_{0}$, one can also construct an invariant
subspace by acting $\eta ^{+}$ on such state. The new degenerate subspace
belongs to the different eigenvalue of $\eta ^{2}$. Taking the same
procedures we have done in $H_{\mathrm{doub}}$, a local imaginary field can
induce a hypercube-like Hamiltonian as in Eq. (\ref{H_eff}) which shares the
same EP with $H_{\mathrm{doub}}$. The order of EP depends on the degeneracy
of eigenenergy of $H_{0}$. This can be demonstrated in Fig. \ref%
{fig_spectrum}. Notice that the system can harbour many EPs with same order
that can be identified by the spin symmetry. This result confirms our
conclusion on the one hand, and tells us that the cooperation between the
local imaginary field and pair fluctuation can accomplish a great task with
little effort by clever maneuvers on the other hand. The dramatic change of
the eigenstates around EP is the key to understand the following interesting
dynamics.

\section{Dynamical preparation of a steady state with ODLRO}

\label{dynamics} In Yang's seminar paper \cite{Yang1989}, the $\eta $%
-pairing states are metastable so that they cannot exist stably in the real
physical system due to destructive interference from the short-range
coherence. In this section, we propose a dynamic scheme to generate a steady
state with ODLRO. The scheme is based on the intriguing features of
high-order EPs under the influence of a local imaginary field. It can be
seen from the previous section that the system spectrum at the EP consists
of many types of high-order EPs. As such the system can be decomposed into
multiple Jordan blocks owing to the existence of various high-order EPs. In
each subspace, an arbitrary initial state will evolve towards the coalescent
state and its probability will increase over time in power law according to
the order of EP \cite{Wang2016,Yang2018}. Then, a natural question arises:
How is the dynamics of such a critical non-Hermitian system? It can be
speculated that for an arbitrary initial state, its time evolution depends
on the interplay among different types of EP. If the evolution time is long
enough, then the highest order EP will determine the final state as the
probability in its subspace grows fastest than those in the other subspaces.
From this point, the steady state can be generated through dynamical
evolution at EP. In the following, we will demonstrate this fascinating
behavior through a critical interacting system and investigate the pair
correlation of the final evolved state.

Let us consider the critical system with the local imaginary field and pair
fluctuation, the Hamiltonian of which can be expressed by setting $j=1$ and $%
\lambda =\gamma $ of Eq. (\ref{HI}). The scheme is that taking the $%
\left\vert N,-N\right\rangle =\left\vert \mathrm{Vac}\right\rangle $ as an
initial state and the final state can be achieved by driving critical
non-Hermitian Hamiltonian. According to the aforementioned statement, the
effective Hamiltonian $H_{\mathrm{doub}}$ is the key to arrive at the
analytical expression of propagator $\mathcal{U=}$ exp$(-iH_{\mathrm{doub}%
}t) $. For simplicity, we first transform $H_{\mathrm{doub}}$ to a standard
Jordan block form\ (block upper triangular form). Notice that $H_{\mathrm{%
doub}}$ is a nilpotent matrix with order $2N+1$ and the geometric
multiplicity of $\left\vert \Phi _{\mathrm{c}}\right\rangle $ is $1$.
Therefore, $2N$ generalized eigenstates should be introduced to complete
this transformation. Such states $\{\left\vert \Phi _{\mathrm{c}%
}^{r_{i}}\right\rangle $, $i=1...2N\}$ can be generated from $\left\vert
\Phi _{\mathrm{c}}\right\rangle $ (see Appendix A for more details).
Performing transformation $A=\left[ \left\vert \Phi _{\mathrm{c}%
}\right\rangle \text{, }\left\vert \Phi _{\mathrm{c}}^{r_{1}}\right\rangle
\text{..., }\left\vert \Phi _{\mathrm{c}}^{r_{2N}}\right\rangle \right] $ on
$H_{\mathrm{doub}}$, we can obtain $H_{\mathrm{doub}}^{\mathrm{s}}=A^{-1}H_{%
\mathrm{doub}}A$ whose matrix element is $\left\langle \overline{N,A_{l}}%
\right\vert H_{\mathrm{doub}}^{\mathrm{s}}\left\vert N,A_{l^{\prime
}}\right\rangle =\delta _{l+1,l^{\prime }}g_{1}/2N$ with $\left\{ \left\vert
N,A_{l}\right\rangle =A^{-1}\left\vert N,l\right\rangle \right\} $ and $%
\left\{ \left\vert \overline{N,A_{l}}\right\rangle =A^{\dagger }\left\vert
N,l\right\rangle \right\} $. Straightforward algebra shows that the
propagator in this new frame can be given as
\begin{equation}
\left\langle \overline{N,A_{l}}\right\vert A^{-1}\mathcal{U}A\left\vert
N,A_{l^{\prime }}\right\rangle =\frac{\left( -it/2N\right) ^{l^{\prime }-l}}{%
\left( l^{\prime }-l\right) !}h\left( l^{\prime }-l\right) ,
\end{equation}%
where $h\left( x\right) $ is a Heaviside step function with the form of $%
h(x)=1$ ($x\geq 0$), and $h(x)=0$ ($x< 0$). Notice that the overall phase $%
e^{-iNU/2}$ is neglected since it does not affect the physical observable.
Considering an arbitrary initial state of this subspace, $%
\sum_{l}c_{l}\left( 0\right) \left\vert N,A_{l}\right\rangle $, the
coefficient $c_{m}\left( t\right) $ of the evolved state can be given as
\begin{equation}
c_{m}\left( t\right) =\sum_{l}\frac{\left( -it/2N\right) ^{l-m}}{\left(
l-m\right) !}h\left( l-m\right) c_{l}\left( 0\right) .
\end{equation}%
It clearly shows that no matter how the initial state is selected, the
coefficient $c_{-N}\left( t\right) $ of evolved state always contains the
highest power of time $t$. Consequently, the component $c_{-N}\left(
t\right) $ of the evolved state overwhelms the other components ensuring the
final state is a coalescent state $\left\vert N,A_{-N}\right\rangle $ under
the Dirac normalization. The different types of the initial state just
determines how the total Dirac probability of the evolved state increases
over time.\ Now we return to the original frame, the final state can be
given as $\left\vert \Phi _{\mathrm{c}}\right\rangle =A\left\vert
N,A_{-N}\right\rangle $, which is the coalescent state of $H_{\mathrm{doub}}$
at EP. According to the Appendix B, the correlation of the expected steady
state is%
\begin{equation}
\left\langle \Phi _{\mathrm{c}}\right\vert \eta _{i}^{+}\eta
_{j}^{-}\left\vert \Phi _{\mathrm{c}}\right\rangle =\left\{
\begin{array}{c}
1/4\text{, for }i\neq j \\
1/2\text{, for }i=j%
\end{array}%
\right. .
\end{equation}%
It has ODLRO and is independent of the relative distance $r=j-i\neq 0$
between the two operators. This intriguing property is in stark difference
to the result obtained in Ref. \cite{Tindall2019}, in which the correlation
of the steady state decay as the increase of $r$. Furthermore, we
investigate the other correlator $\left\langle \Phi _{\mathrm{c}}\right\vert
\eta _{i}^{+}\left\vert \Phi _{\mathrm{c}}\right\rangle $ that is shown to
be zero in the recent experimental and theoretical study \cite%
{Tindall2019,Kaneko2019}. Straightforward algebra shows that $\left\langle
\Phi _{\mathrm{c}}\right\vert \eta _{i}^{+}\left\vert \Phi _{\mathrm{c}%
}\right\rangle =-i/2$. Such quantity is reminiscent of the order parameter $%
\langle b_{i}^{\dagger }\rangle $ ($b_{i}^{\dagger }$ represents the boson
creation operator) that describes the quantum phase transition from the Mott
insulator to superfluid in the Bose-Hubbard model. The nonzero value may
imply some important physical property and give further insight into the
non-Hermitian Hubbard model.
\begin{figure}[tbp]
\centering
\includegraphics[width=0.48\textwidth]{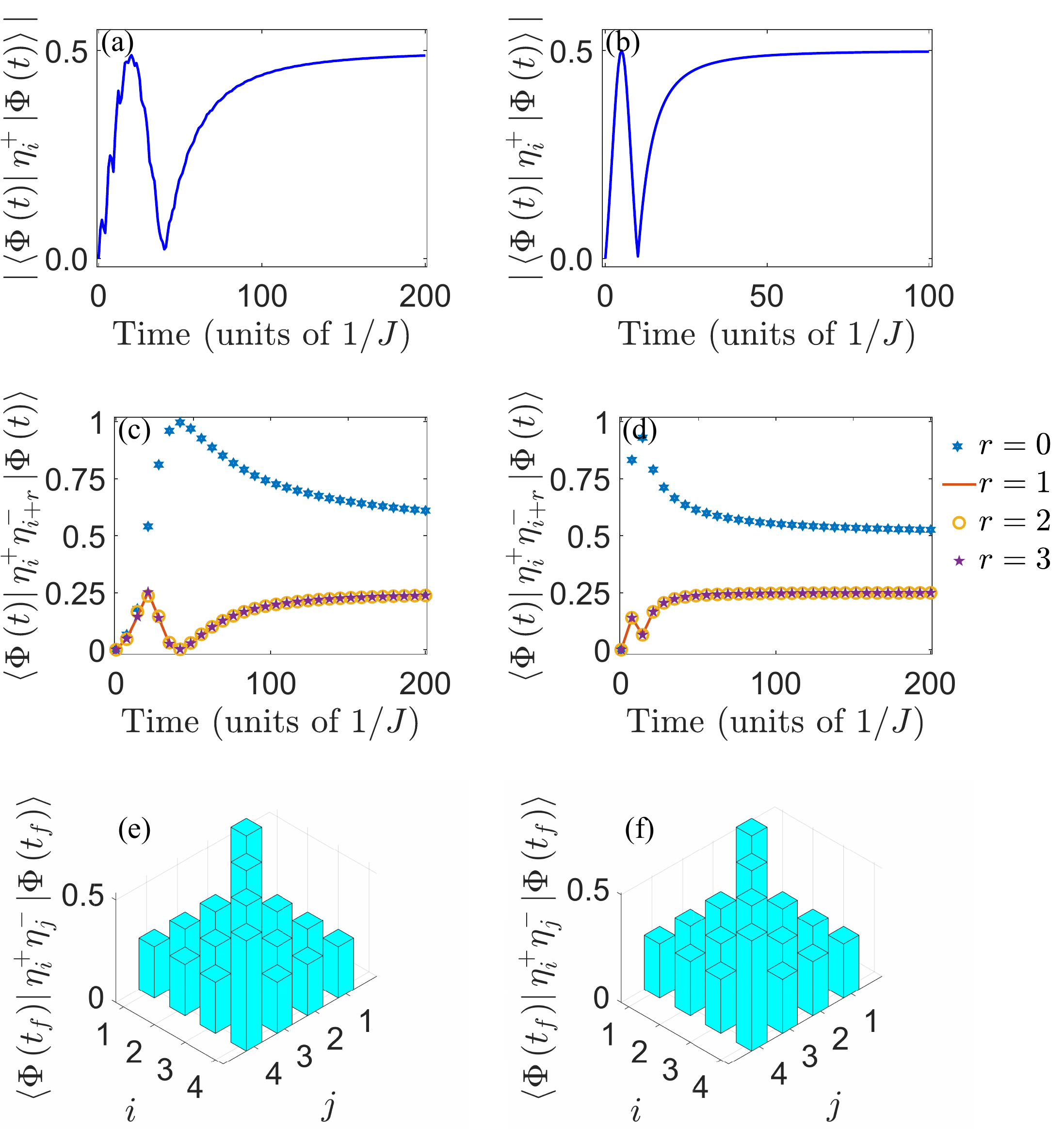}
\caption{(a)-(d) Evolution of the correlators $|\left\langle \Phi \left(
t\right) \right\vert \protect\eta _{i}^{+}\left\vert \Phi \left( t\right)
\right\rangle |$ and $\left\langle \Phi \left( t\right) \right\vert \protect%
\eta _{i}^{+}\protect\eta _{i+r}^{-}\left\vert \Phi \left( t\right)
\right\rangle $, averaged over all sites for the $4$ site Hubbard model. The
initial state is prepared in vacuum state $\left\vert \mathrm{Vac}%
\right\rangle $ of $H_{0}$ with interaction $U=2t$, and then it is driven by
the system with the local imaginary field $g_{1}=g=0.2t$ for (a) and (c),
and homogeneous dissipation $g_{j}=g=0.2t$ $(j=1...,N)$ for (b) and (d),
respectively. Notice that $H_{I}$ is at EP such that $\protect\lambda /%
\protect\gamma =1$. (e)-(f) The correlation values of steady state for
different relative distance ($\left\langle \Psi \left( t_{f}\right)
\right\vert \protect\eta _{i}^{+}\protect\eta _{j}^{-}\left\vert \Psi \left(
t_{f}\right) \right\rangle $) at relaxation time $t_{f}=400t$ for (e) and $%
t_{f}=100t$ for (f). It is shown that $\left\langle \Psi \left( t_{f}\right)
\right\vert \protect\eta _{i}^{+}\protect\eta _{j}^{-}\left\vert \Psi \left(
t_{f}\right) \right\rangle =1/4$ for $i\neq j$ and $\left\langle \Psi \left(
t_{f}\right) \right\vert \protect\eta _{i}^{+}\protect\eta %
_{j}^{-}\left\vert \Psi \left( t_{f}\right) \right\rangle =1/2$ for $i=j$,
which confirms the understanding in the main text.}
\label{fig_dynamics}
\end{figure}

To check this understanding, we perform a numerical simulation and present
the results in Fig. \ref{fig_dynamics} with an initial state being $%
\left\vert \Psi \left( 0\right) \right\rangle =\left\vert N,-N\right\rangle $%
. We inspect the time dependence of two correlators $|\left\langle \Psi
\left( t\right) \right\vert \eta _{i}^{+}\left\vert \Psi \left( t\right)
\right\rangle |$ and $\left\langle \Psi \left( t\right) \right\vert \eta
_{i}^{+}\eta _{j}^{-}\left\vert \Psi \left( t\right) \right\rangle $. It can
be shown that the final values of the two correlators are $1/2$, and $1/4$,
which agree with our analytical results. The number of local imaginary
fields only changes the relaxation time $t_{r}$ and does not change the
final value of the correlator. The systems that realize the proposed dynamic
scheme are all with the reach of ongoing experiments. Hence, such a scheme
offers a unique perspective to generate a steady superconducting-like state
in a variety of materials.

\section{Summary}

\label{summary} In summary, we have investigated some general aspects of the
non-Hermitian Hubbard model. We have shown that the presence of the local
imaginary field and pair fluctuation can drastically change its microscopic
behavior and manifest a variety of collective and cooperative phenomena at
the macroscopic level. Specifically, the whole real spectrum of such a
non-Hermitian interacting system can be achieved in a wide range of
parameters even though a local imaginary field is applied. At EP, the
interplay between the local imaginary field and on-site pair fluctuation can
lead to a coalescent state with the geometric multiplicity of $1$, which is
protected by the on-site pair interaction. $\eta $-pairing symmetry plays
the key role to understand the formation of the high-order EP. Comparing
with the other schemes in Refs. \cite{Tindall2019,Kaneko2019} that produce a
superconducting-like state, the critical strongly correlated system in our
scheme favors superconductivity on a long time scale. The corresponding
steady state with ODLRO can be generated through time evolution from an
arbitrary initial state. The realization of this scheme does not depend on
the location of the local imaginary filed and pair fluctuation. Hence, this
scheme does open up the possibility of generating a steady
superconducting-like state with a long coherence time in a variety of
experimental platforms. \acknowledgments We acknowledge the support of the
National Natural Science Foundation of China (Grants No. 11975166, and No.
11874225). X.Z.Z. is also supported by the Program for Innovative Research
in University of Tianjin (Grant No. TD13-5077).

\appendix
\label{appendix}

\section{the derivation of the generalized eigenstates $\{\left\vert \Phi _{%
\mathrm{c}}^{r_{i}}\right\rangle \}$}

In this subsection, we demonstrate how to generate generalized eigenstates $%
\{\left\vert \Phi _{\mathrm{c}}^{r_{i}}\right\rangle \}$. When $\gamma
=\lambda $, the effective Hamiltonian $H_{\mathrm{doub}}$ is a nilpotent
matrix with order $2N+1$ such that $\left( H_{\mathrm{doub}}\right)
^{2N+1}=0 $ and the coalescent eigenstate $\left\vert \Phi _{\mathrm{c}%
}\right\rangle $ has the geometric multiplicity of $1$. Therefore, one
should introduce $2N$ generalized eigenstates $\{\left\vert \Phi _{\mathrm{c}%
}^{r_{i}}\right\rangle $ to perform the transformation $A=\left[ \left\vert
\Phi _{\mathrm{c}}\right\rangle \text{, }\left\vert \Phi _{\mathrm{c}%
}^{r_{1}}\right\rangle \text{..., }\left\vert \Phi _{\mathrm{c}%
}^{r_{2N}}\right\rangle \right] $, which transform the Hamiltonian $H_{%
\mathrm{doub}}$ to a standard Jordan block $H_{\mathrm{doub}}^{\mathrm{s}%
}=A^{-1}H_{\mathrm{doub}}A$ with upper triangular form. The generalized
eigenstates $\{\left\vert \Phi _{\mathrm{c}}^{r_{i}}\right\rangle \}$ can be
generated by $\left\vert \Phi _{\mathrm{c}}^{r_{i}}\right\rangle $ following
the steps below:

First, we consider the relation%
\begin{equation}
H_{\mathrm{doub}}A=AH_{\mathrm{doub}}^{\mathrm{s}},
\end{equation}%
where $H_{\mathrm{doub}}$ and $H_{\mathrm{doub}}^{\mathrm{s}}$ are in the
matrix form %\begin{widetext}
\begin{equation}
H_{\mathrm{doub}}=\frac{\lambda G}{4N}\left[
\begin{array}{cccc}
-2iN & J_{-N} &  &  \\
J_{-N} & \ddots  & \ddots  &  \\
& \ddots  & \ddots  & J_{N-1} \\
&  & J_{N-1} & 2iN%
\end{array}%
\right] ,
\end{equation}%
%
%
%
%
%
%
%
%
%
%
%
%
%
%
%
%\end{widetext}
and%
\begin{equation}
H_{\mathrm{doub}}^{\mathrm{s}}=\frac{\lambda G}{2N}\left[
\begin{array}{cccc}
& 1 &  &  \\
&  & \ddots  &  \\
&  &  & 1 \\
&  &  &
\end{array}%
\right] .
\end{equation}%
Here we omit the on-site interaction term for convenience. Then the
generalized eigenstates can be obtained by applying $H_{\mathrm{doub}}$ to
coalescent state step by step,
\begin{eqnarray}
H_{\mathrm{doub}}\left\vert \Phi _{\mathrm{c}}\right\rangle  &=&0\left\vert
\Phi _{\mathrm{c}}\right\rangle ,  \label{ir_1} \\
H_{\mathrm{doub}}\left\vert \Phi _{\mathrm{c}}^{r_{1}}\right\rangle
&=&\left\vert \Phi _{\mathrm{c}}\right\rangle ,  \label{ir_2} \\
&&\cdots   \notag \\
H_{\mathrm{doub}}\left\vert \Phi _{\mathrm{c}}^{r_{i+1}}\right\rangle
&=&\left\vert \Phi _{\mathrm{c}}^{r_{i}}\right\rangle ,  \label{ir_3} \\
H_{\mathrm{doub}}\left\vert \Phi _{\mathrm{c}}^{r_{2N}}\right\rangle
&=&\left\vert \Phi _{\mathrm{c}}^{r_{2N-1}}\right\rangle .  \label{ir_4}
\end{eqnarray}%
In the following, we take $N=1$ as an example to give the concrete
expression of $A$. Starting from $\left\vert \Phi _{\mathrm{c}}\right\rangle
=\left[ 1,\text{ }\sqrt{2}i,\text{ }-1\right] ^{T}/2$, we can obtain $%
\left\vert \Phi _{\mathrm{c}}^{r_{1}}\right\rangle =\left[ i,\text{ }\sqrt{2}%
/2,\text{ }0\right] ^{T}$ according to Eq. (\ref{ir_2}). Obviously, the
selection of $\left\vert \Phi _{\mathrm{c}}^{r_{1}}\right\rangle $ is not
unique. With the aid of relation $H_{\mathrm{doub}}\left\vert \Phi _{\mathrm{%
c}}^{r_{2}}\right\rangle =\left\vert \Phi _{\mathrm{c}}^{r_{1}}\right\rangle
$, the transformation matrix of $A$ can be given as
\begin{equation}
A=\frac{1}{2}\left(
\begin{array}{ccc}
1 & 2i & -2 \\
\sqrt{2}i & -\sqrt{2} & 0 \\
-1 & 0 & 0%
\end{array}%
\right) .
\end{equation}%
One can check that $H_{\mathrm{doub}}^{\mathrm{s}}=A^{-1}H_{\mathrm{doub}}A$%
. Here we emphasize that population in $\left\vert \Phi _{\mathrm{c}%
}^{r_{i+1}}\right\rangle $ is transferred to $\left\vert \Phi _{\mathrm{c}%
}^{r_{i}}\right\rangle $ during the time evolution owing to the fact that $%
H_{\mathrm{doub}}\left\vert \Phi _{\mathrm{c}}^{r_{i+1}}\right\rangle
=\left\vert \Phi _{\mathrm{c}}^{r_{i}}\right\rangle $. After the relaxation
time, all the initial states will evolve to the final coalescent state. This
mechanism serves as the building block to generate a steady state with ODLRO.

\section{derivation of two correlators}

In this subsection, we concentrate on the correlator $\left\langle \Phi _{%
\mathrm{c}}\right\vert \eta _{i}^{+}\eta _{j}^{-}\left\vert \Phi _{\mathrm{c}%
}\right\rangle $. Before proceeding the calculation, we have known that the
final evolved state $\left\vert \Phi _{\mathrm{c}}\right\rangle $ is a
high-order coalescent state with geometric multiplicity being $1$. The wave
function can be given as
\begin{equation}
\langle N,l\left\vert \Phi _{\mathrm{c}}\right\rangle =(\frac{i}{2})^{N}%
\sqrt{C_{2N}^{N+l}}\left( i\right) ^{l},
\end{equation}%
where $\left\vert N,l\right\rangle =\Omega ^{-1}\left( \eta ^{+}\right)
^{N+l}\left\vert \mathrm{Vac}\right\rangle $ with $\Omega =\sqrt{C_{2N}^{N+l}%
}$. Straightforward algebra shows that
\begin{equation}
\eta _{j}^{-}\left\vert \Phi _{\mathrm{c}}\right\rangle =(\frac{i}{2}%
)^{N}\sum_{l}\left( i\right) ^{l}(\sum_{n\neq j}\eta
_{n}^{+})^{N+l}\left\vert \mathrm{Vac}\right\rangle .
\end{equation}%
Combining with
\begin{equation}
\eta _{i}^{-}\left\vert \Phi _{\mathrm{c}}\right\rangle =(\frac{i}{2}%
)^{N}\sum_{l}\left( i\right) ^{l}(\sum_{n\neq i}\eta
_{n}^{+})^{N+l}\left\vert \mathrm{Vac}\right\rangle ,
\end{equation}%
one can readily obtain
\begin{equation}
\left\langle \Phi _{\mathrm{c}}\right\vert \eta _{i}^{+}\eta
_{j}^{-}\left\vert \Phi _{\mathrm{c}}\right\rangle =\left( 4\right)
^{-N}\sum_{l=-N}^{N-2}C_{2N-2}^{N+l}=\frac{1}{4}.
\end{equation}%
Interestingly, the correlation function of such a steady state is irrelevant
to the relative distance $r=j-i$ between the two operators. This feature may
facilitate future experiment in generating superconduting-like state.

On the other hand, we investigate the correlator $\left\langle \Phi _{%
\mathrm{c}}\right\vert \eta _{i}^{+}\left\vert \Phi _{\mathrm{c}%
}\right\rangle $. With the same spirit, one can give the expression
\begin{equation}
\eta _{i}^{+}\left\vert \Phi _{\mathrm{c}}\right\rangle =(\frac{i}{2}%
)^{N}\sum_{l}\left( i\right) ^{l}(\sum_{n\neq i}\eta _{n}^{+})^{N+l}\eta
_{i}^{+}\left\vert \mathrm{Vac}\right\rangle ,
\end{equation}%
which yields the result
\begin{equation}
\left\langle \Phi _{\mathrm{c}}\right\vert \eta _{i}^{+}\left\vert \Phi _{%
\mathrm{c}}\right\rangle =-i2^{-N}\sum_{l=-N}^{N-1}C_{2N-1}^{N+l}=-\frac{i}{2%
}.
\end{equation}

%\bibliography{reference2}

\end{document}